\documentclass[11pt]{article}
\usepackage{epsfig}

\newcommand{\BABARPubYear}    {02}

\newcommand{\BABARProcNumber} {23}
\newcommand{\SLACPubNumber} {9219}
\newcommand{\LANLNumber} {0205056}

\input pubboard/babarsym

\long\def\inst#1{\par\nobreak\kern 4pt\nobreak
    {\it #1}\par\vskip 10pt plus 3pt minus 3pt}

\begin{document}
{\pagestyle{empty}

\begin{flushright}
SLAC-PUB-\SLACPubNumber \\
\babar-PROC-\BABARPubYear/\BABARProcNumber \\
hep-ex/\LANLNumber \\
May, 2002 \\
\end{flushright}

\par\vskip 4cm

\begin{center}
\Large \bf Rare $B$ Decays at \Lbabar\
\end{center}
\bigskip

\begin{center}
\large 
Dieter Best\\
University of California, Irvine \\
Department of Physics and Astronomy \\
Irvine, CA 92697-4575 \\
(on behalf of the \lbabar\ Collaboration)
\end{center}
\bigskip \bigskip

\begin{center}
\large \bf Abstract
\end{center}
We report recent results in the search for the rare $B$ meson decays $B\rightarrow\rho\gamma$ 
and $B^0\rightarrow\pi^0\pi^0$. These results are based on 56.4 fb$^{-1}$ 
collected by the BaBar Collaboration at the SLAC PEP-II $e^+e^-$ $B$ Factory.
We set new  90\% confidence level upper limits 
${\cal B}(B^0\rightarrow\rho^0\gamma) < 1.5 \times 10^{-6}$,
${\cal B}(B^+\rightarrow\rho^+\gamma) < 2.8 \times 10^{-6}$, and 
${\cal B}(B^0\rightarrow\pi^0\pi^0)   < 3.4 \times 10^{-6}$.

\vfill
\begin{center}
Invited talk presented at the XXXVIIth Rencontres de Moriond 
on QCD and Hadronic Interactions, \\
3/16/2002---3/23/2002, Les Arcs, France
\end{center}

\vspace{1.0cm}
\begin{center}
{\em Stanford Linear Accelerator Center, Stanford University, 
Stanford, CA 94309} \\ \vspace{0.1cm}\hrule\vspace{0.1cm}
Work supported in part by Department of Energy contract DE-AC03-76SF00515.
\end{center}

\def\Vud{V_{ud}}
\def\Vudabs{\vert V_{ud} \vert}
\def\Vus{V_{us}}
\def\Vusabs{\vert V_{us} \vert}
\def\Vub{V_{ub}}
\def\Vubabs{\vert V_{ub}\vert}
\def\Vcd{V_{cd}}
\def\Vcdabs{\vert V_{cd} \vert}
\def\Vcs{V_{cs}}
\def\Vcsabs{\vert V_{cs} \vert}
\def\Vcb{V_{cb}}
\def\Vcbabs{\vert V_{cb} \vert}
\def\Vtd{V_{td}}
\def\Vtdabs{\vert V_{td} \vert}
\def\Vts{V_{ts}}
\def\Vtsabs{\vert V_{ts} \vert}
\def\Vtb{V_{tb}}

\section{The BaBar Detector}

The results presented in this paper are based on an integrated luminosity of 
56.4 fb$^{-1}$ collected on the $\Upsilon(4S)$ resonance with the BaBar detector \cite{BaBar} 
at the PEP-II asymmetric $e^+e^-$ collider of the Stanford Linear Accelerator Center. 

Charged particles are detected and their momenta measured by a combination of a 
5 double--sided layer silicon vertex tracker
and a 40--layer drift chamber, photons are detected by a CsI electromagnetic calorimeter.
These detectors operate inside  a 1.5 T solenoidal field.
 
Charged particle identification is achieved by the average energy loss in the tracking devices 
and by a unique, internally reflecting ring imaging Cherenkov detector. 

\section{$B$ Decay Reconstruction}

The $B$ meson candidates are kinematically identified using two independent variables. 
$\Delta E = E^*_B - E^*_{Beam}$ is peaked at zero for signal since the $B$ mesons are
produced via $e^+e^- \rightarrow \Upsilon(4S) \rightarrow B\overline B$ and therefore
the energy of the $B$ meson in the  $\Upsilon(4S)$ rest frame is the beam energy $E^*_{Beam}$.
$m_{ES} = \sqrt{ E^{*2}_{Beam} - {\mathbf p}_B^{*2} }$ is a measure of the $B$ meson mass where
we have substituted $ E^*_B $ by the beam energy $ E^*_{Beam}$ which is known with better
precision than $E^*_B$. ${\bf p}_B^{*}$ is the momentum of the  $B$ meson candidate in the $\Upsilon(4S)$ 
rest frame calculated from the measured momenta of the decay products.

Rare $B$ decay modes suffer from large backgrounds due to random combinations of tracks
produced in the light quark--antiquark continuum. The distinguishing feature of such 
backgrounds is their characteristic event shape resulting from the two--jet production
mechanism. 
A quantity that characterizes the event shape is the angle between the thrust
axis of the $B$ candidate and the thrust axis of the rest of the event where the thrust 
axis is defined as the axis that maximizes the sum of the magnitudes of the longitudinal
momenta. This angle is small for continuum events, where the $B$ candidate daughters
tend to lie in the $q\overline q$ jets, and uniformly distributed for true $B\overline B$ 
events.

We further suppress background using a Fisher discriminant constructed as an optimized linear combination 
of the scalar sum of the center--of--mass
momenta of all charged tracks and photons (excluding the $B$ candidate decay products) flowing into 
9  concentric cones centered on the thrust axis of the $B$ candidate.
The more spherical the event, the lower the value of the Fisher discriminant.

All analyses have been performed as {\em blind analyses}, {\em i.e.}, the region in $\Delta E$ and $m_{ES}$
where the signal is expected is concealed until all the selection criteria are determined either from
Monte Carlo events, data sidebands in the $\Delta E$--$m_{ES}$ plane (region outside the signal region) or 
data control samples.

Yields are extracted using extended maximum likelihood fits to the $\Delta E$, $m_{ES}$, and Fisher discriminant
(for $B^0\rightarrow \pi^0\pi^0$) or resonance mass distribution (for $B\rightarrow\rho\gamma$) for signal
and background.

In the following charge conjugate modes are implied throughout.

\section{A Search for $B\rightarrow\rho\gamma$}

The measurement of the branching fraction for $B \rightarrow \rho \gamma$ is mainly aiming at
the determination of the quark--mixing matrix element $|V_{td}|$. 
The measurement of $|V_{td}|$ is also the main motivation for performing $B^0-\overline{B}^0$ mixing measurements.
Both are sensitive to new physics contributions in their loop and box diagrams.

Assuming that the short--distance  contribution
of the magnetic moment operator is dominating these transitions,
one derives 
\begin{equation}
  \label{ratio1}
  \frac{\Gamma(B \to \rho/\omega \gamma)}{\Gamma(B \to K^* \gamma)}
  = \left( \frac{\Vtdabs}{\Vtsabs}\right)^2 \xi \Omega ~,
\end{equation}
where $\xi$ takes into account the decay form factors and 
$\Omega$ the different phase space factors \cite{ABS93}.
With $|V_{ts}| \approx |V_{cb}|$ a precise measurement of the ratio
of branching fractions provides therefore a strong constraint on $|V_{td}|$.

With the assumption of isospin invariance one also expects
$\Gamma(B^{+} \to \rho^+ \gamma)=2 ~\Gamma(B^{0} \to \rho^0  \gamma)$,
if we assume in addition SU(3) symmetry we obtain 
$\Gamma(B^{0} \to \rho^0  \gamma) = \Gamma (B^{0} \to \omega  \gamma)$.
The measurement of the branching fractions alone provides therefore a model
independent way of testing to what extent the short distance contributions
are dominating these decays.

The relations above have been used to convert the experimental upper bound on the ratio of the
exclusive radiative $B$ decays \cite{Patterson}
$ {\cal B}(B \to \rho/\omega \gamma)/{\cal B}(B \to K^* \gamma) < 0.34 $ (90\% confidence level),
into a bound on $\Vtdabs/\Vtsabs$, $<$ 0.64 -- 0.76,
depending on the estimate of the SU(3)--breaking parameters in the short distance
contribution \cite{Narison}.
While this bound is at present not competitive
with the corresponding bound from the unitarity of the quark--mixing matrix \cite{PDG} and
from the fits of the quark--mixing matrix elements \cite{AL94}, which yield
$\Vtdabs/\Vtsabs < 0.36$,
one anticipates that the increased sensitivity in the
radiative $B$ decay modes at the high luminosity $B$ factories will allow to test
these relationships quantitatively.


Currently no measurements exist for  the branching fraction of the decay $B \rightarrow \rho \gamma$, however,
90\% confidence level upper limits have been obtained by the CLEO~\cite{CLEOrhoGamma} 
and Belle collaborations~\cite{BelleRhoGamma}.
CLEO obtains an upper limit of 17$\times$10$^{-6}$ for the neutral and 13$\times$10$^{-6}$ for the charged mode
based on 9.7$\times$10$^{6}$ $B\overline B$ events. 
Belle obtains an upper limit of 10.6$\times$10$^{-6}$ for the neutral and 9.9$\times$10$^{-6}$ for the charged mode
based on 11$\times$10$^{6}$ $B\overline B$ events.


\begin{figure}[ttt]
  \epsfig{file=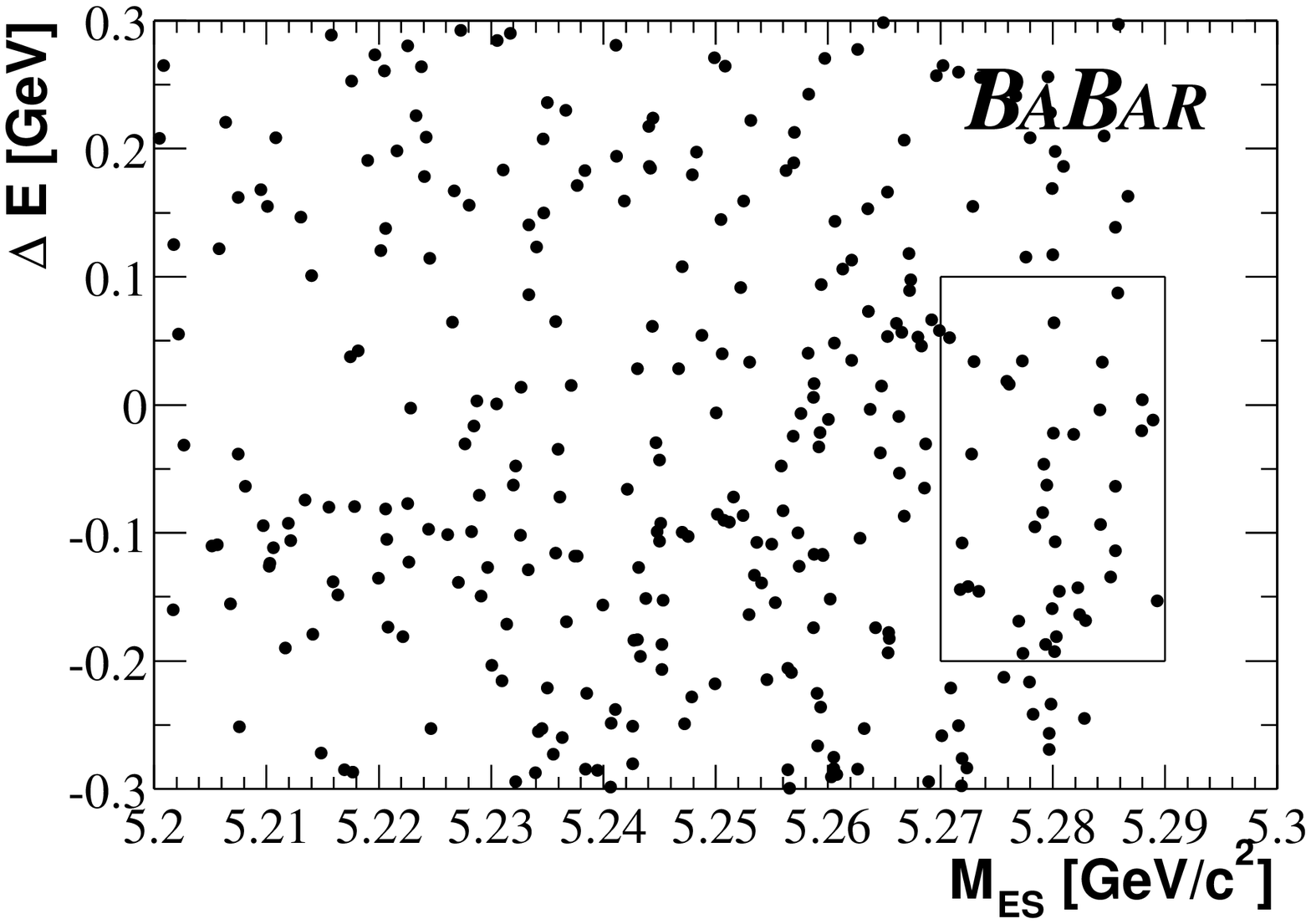,width=0.5\linewidth}\epsfig{file=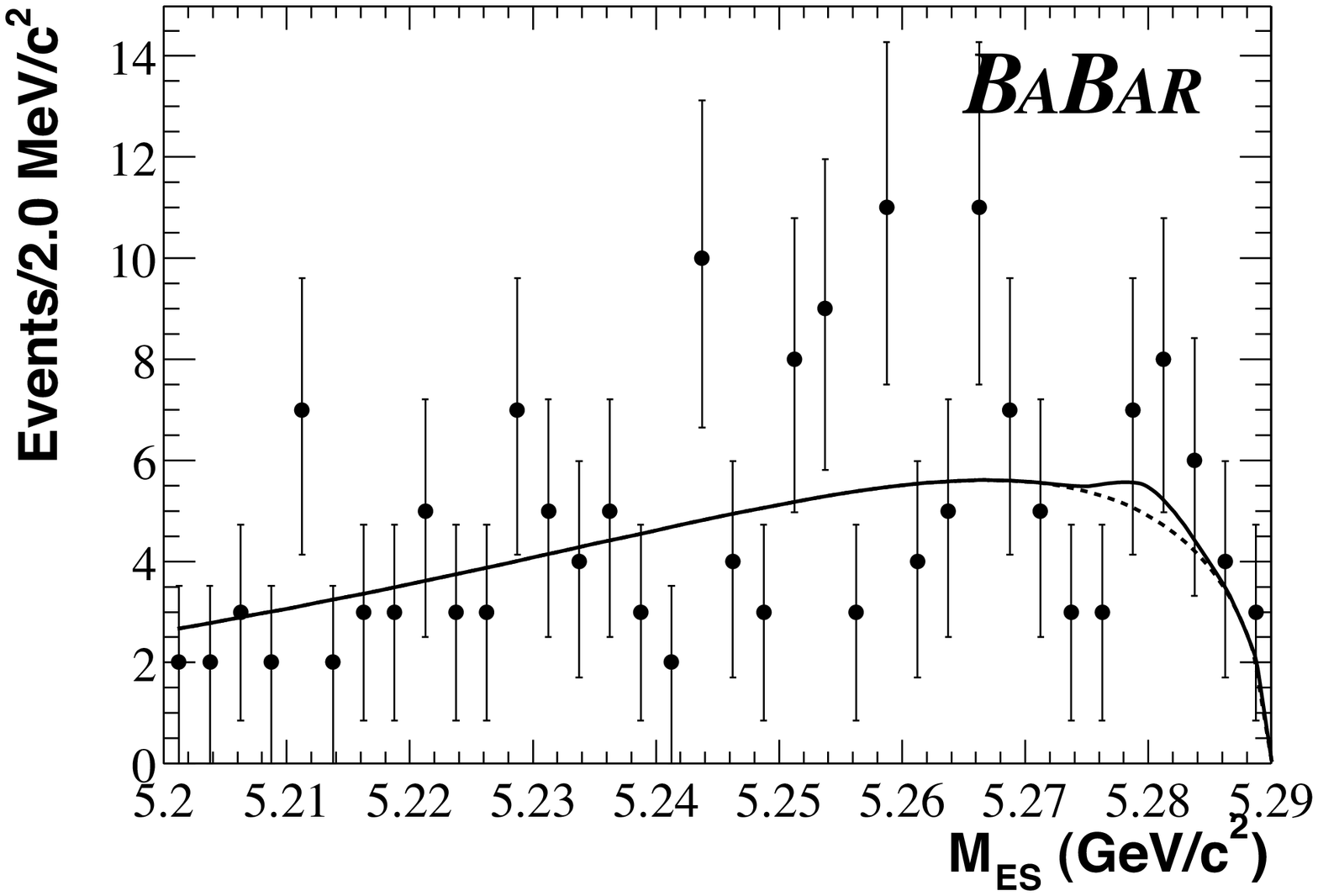,width=0.5\linewidth}
  \caption{Left: $\Delta E$ vs $m_{ES}$ distribution for $B^0 \rightarrow \rho^0 \gamma$. The box outlines the
    signal region which was blinded during the analysis.
    Right: $m_{ES}$ distribution for $B^+ \rightarrow \rho^+ \gamma$. The result of the likelihood fit is 
    overlaid as a solid line, the dashed line represents the background contribution.}
  \label{fig:rhoGamma}
\end{figure}

Figure \ref{fig:rhoGamma} shows our results for $B\rightarrow\rho\gamma$ after unblinding.
On the left is shown the $\Delta E$ vs $m_{ES}$ distribution for $B^0 \rightarrow \rho^0 \gamma$. 
The box outlines the signal region which was blinded during the analysis.
On the right is shown the $m_{ES}$ distribution for $B^+ \rightarrow \rho^+ \gamma$. The result of the likelihood fit is 
overlaid as a solid line, the dashed line represents the background contribution.
The extended maximum likelihood fit yields 3.1 $\pm$ 4.2 events for $B^0\rightarrow\rho^0\gamma$ and
4.6 $\pm$ 5.8 events for  $B^+ \rightarrow \rho^+ \gamma$. We observe no signal and place the following 
90\% confidence level upper limits on the branching fractions:
$$
 {\cal B}(B^0\rightarrow\rho^0\gamma) < 1.5 \times 10^{-6},  \;\;\;\;\;\;\;\;
 {\cal B}(B^+\rightarrow\rho^+\gamma) < 2.8 \times 10^{-6}. 
$$
This represents an improvement of an order of magnitude over previous measurements and reaches the
range of theoretical predictions.
The systematic uncertainty on the branching fraction is about 15\% for both modes, where the largest 
contribution comes from the assumption of the number of $\rho$ mesons in the background.

\begin{figure}[ttt]
\epsfig{file=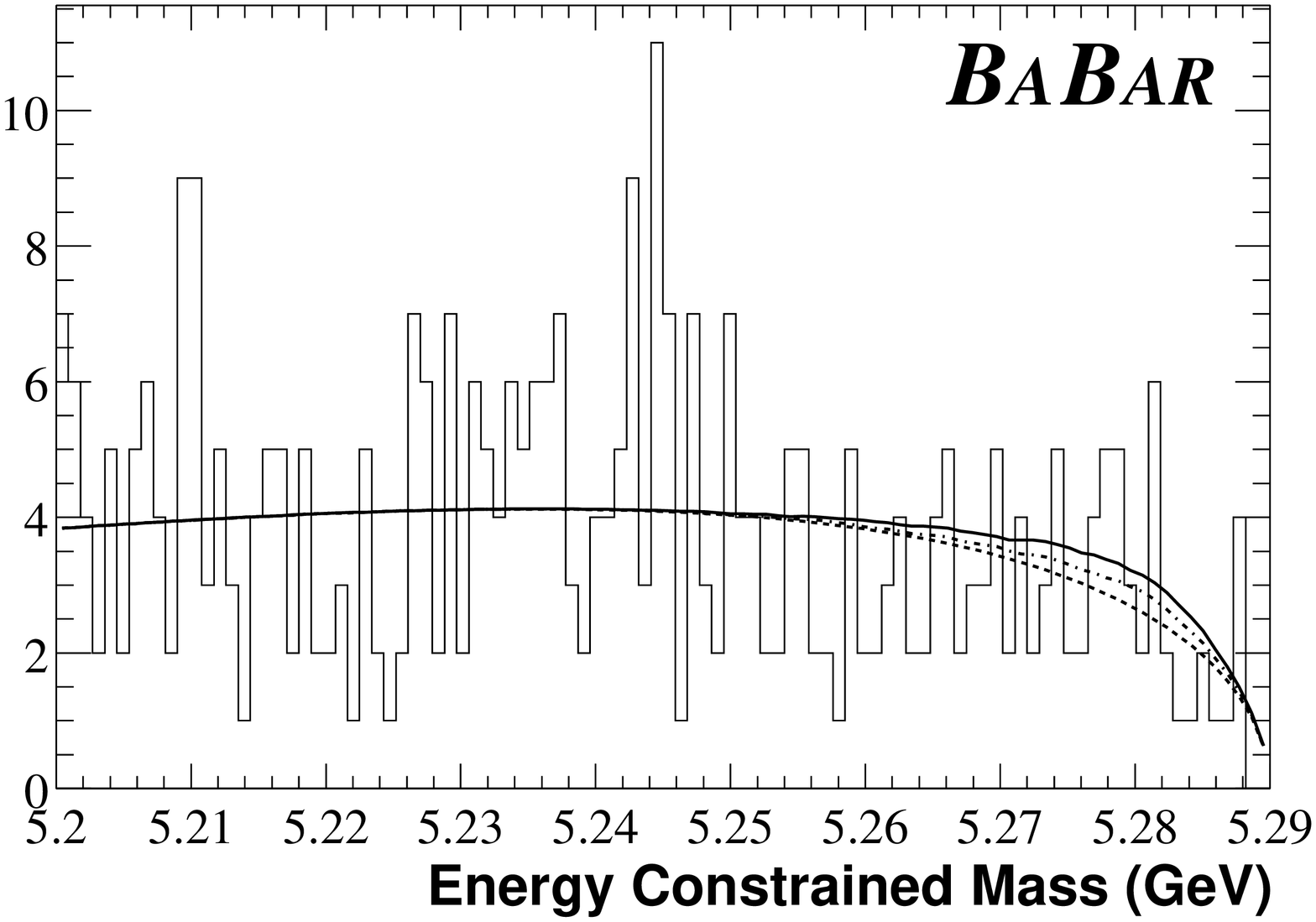,angle=-90.0,width=0.5\linewidth}\epsfig{file=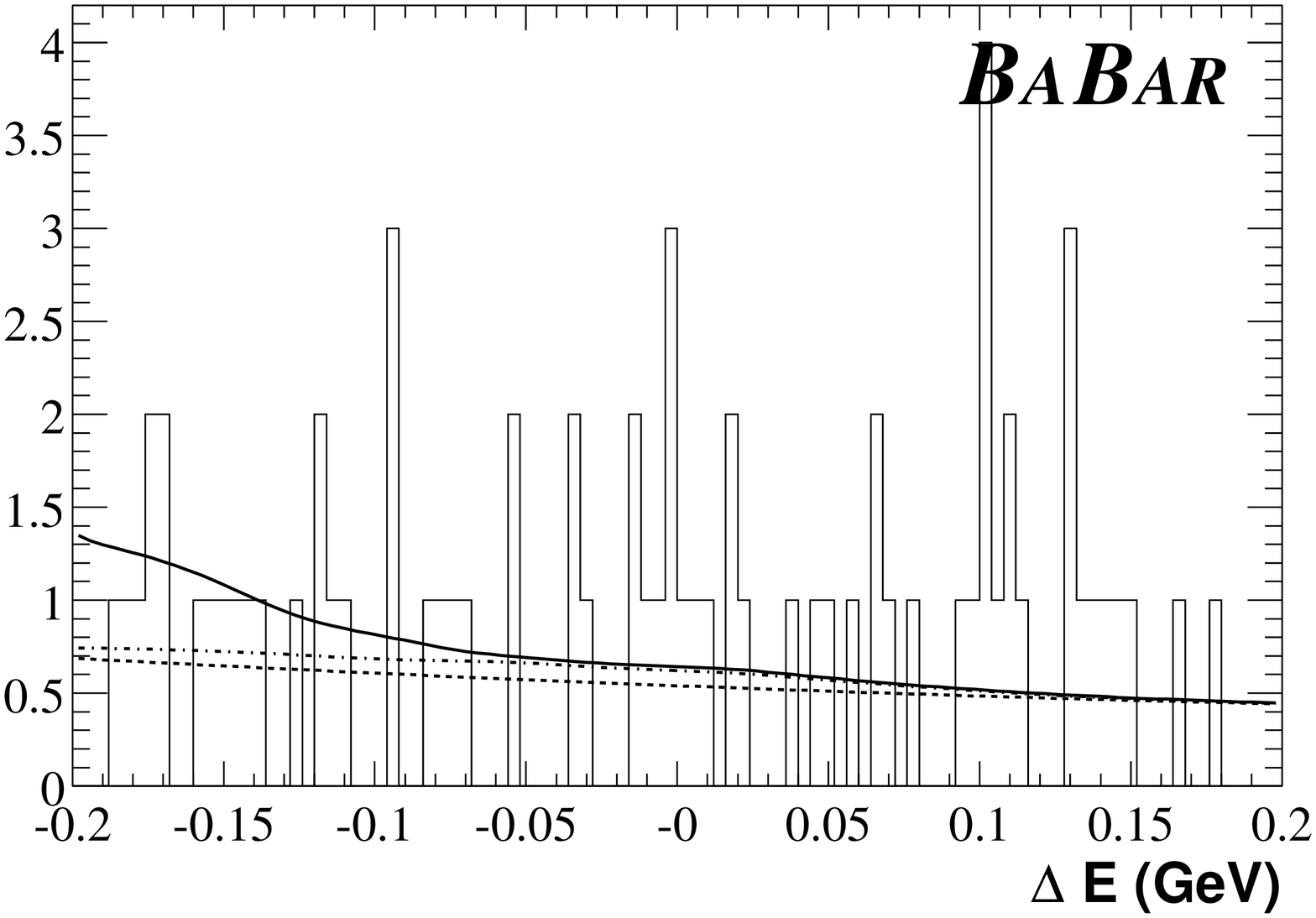,angle=-90.0,width=0.5\linewidth}
\caption{The $m_{ES}$ (left) and $\Delta E$ (right) distributions for events 
  passing the cut and count selection except for the cut on $m_{ES}$ itself. 
  The likelihood fit is overlaid with the correct scaling. The solid line 
  is the overall probability density distribution function (PDF), the dashed line represents the continuum PDF while 
  the dashed-dotted line represents the continuum and $B^0\rightarrow\pi^0\pi^0$ 
  PDFs combined.}
\label{fig:pi0pi0}
\end{figure}

\section{A Search for $B^0\rightarrow \pi^0\pi^0$}

The branching fraction for the decay $B^0\rightarrow \pi^0\pi^0$ is interesting in the
context of the determination of the angle $\alpha$ in the unitarity triangle. 
In the absence of penguin contributions, the asymmetry in $B^0\rightarrow \pi^+\pi^-$
measures $\sin(2\alpha)$. 
An isospin analysis can be used to eliminate the penguin pollution in this case \cite{Gronau}.
However, this analysis requires both the measurement of $B^0\rightarrow \pi^0\pi^0$
and $\overline B^0 \rightarrow \pi^0\pi^0$, and therefore, although theoretically clean, this analysis
is undermined by the small branching fraction of the decay $B^0\rightarrow \pi^0\pi^0$.
Theoretical predictions \cite{Zhou} are as high as 4.6$\times$10$^{-6}$, 
some references \cite{Ali} give limits in the range 10$^{-7}$ -- 10$^{-6}$.

However, measuring a CP averaged decay rate for $B^0\rightarrow \pi^0\pi^0$ is still 
interesting for $\sin(2\alpha)$. The phase angle obtained through the analysis of 
$B^0\rightarrow \pi^+\pi^-$ decays gives only an effective parameter $\alpha_{eff}$ 
which is dependent on  $\alpha$, strong phases and the ratio of penguin to tree amplitudes.
One can bound  $\alpha$ via\cite{Quinn}
$$
\sin^2(\alpha_{eff}-\alpha) = 
\frac{\langle{\cal B}(B^0\rightarrow \pi^0\pi^0)\rangle_{CP}}{{\cal B}(B^+\rightarrow \pi^+\pi^0)}
$$ 
where $\langle{\cal B}(B^0\rightarrow \pi^0\pi^0)\rangle_{CP} = \frac{1}{2} \left [ {\cal B}(B^0\rightarrow \pi^0\pi^0) 
+  {\cal B}(\overline B^0\rightarrow \pi^0\pi^0) \right ]$.

Recent results from CLEO~\cite{CLEO} and Belle~\cite{Belle} have hinted at the possibility of a 
branching fraction of the order of 2$\times$10$^{-6}$. 
CLEO measures a limit of 5.7$\times$10$^{-6}$ based on  9.67$\times$10$^{6}$ $B\overline B$ events,
Belle  5.6$\times$10$^{-6}$  based on 31.7$\times$10$^{6}$ $B\overline B$ events.

Figure \ref{fig:pi0pi0} shows projections on $m_{ES}$ (left) and $\Delta E$ (right) after unblinding
with the likelihood function superimposed. 
The solid line is the overall probability density distribution function (PDF), 
the dashed line represents the continuum PDF while 
the dashed--dotted line represents the continuum and $B^0\rightarrow\pi^0\pi^0$ 
PDFs combined.
There is no evidence for observation of a signal for $B^0\rightarrow\pi^0\pi^0$ yet. We place a 90\%
confidence level upper limit on the branching fraction of 
$$
{\cal B}(B^0\rightarrow\pi^0\pi^0) < 3.4 \times 10^{-6}.
$$
This upper limit is better than that of previous searches.
The largest two systematic uncertainties arise from the parameterization of the continuum background and
the Fisher discriminant distribution, the third largest from the assumption for the
$B^+\rightarrow\rho^+\pi^0$ background contribution.

\end{document}